\documentclass[twocolumn,showpacs,preprintnumbers,amsmath,amssymb]{revtex4}

\usepackage{graphicx}
\usepackage{dcolumn}
\usepackage{bm}

\begin{document}
\title{A Mutual Selection Model for Weighted Networks}
\author{Wen-Xu Wang}
\author{Bo Hu}
\author{Tao Zhou}
\author{Bing-Hong Wang}
\email{bhwang@ustc.edu.cn,Fax:+86-551-3603574}
\author{Yan-Bo Xie}
\affiliation{%
Nonlinear Science Center and Department of Modern Physics,\\
University of Science and Technology of China, Hefei, 230026, P.R.
China
}%

\date{\today}

\begin{abstract}
For most networks, the connection between two nodes is the result
of their mutual affinity and attachment. In this paper, we propose
a mutual selection model to characterize the weighted networks. By
introducing a general mechanism of mutual selection, the model can
produce power-law distributions of degree, weight and strength, as
confirmed in many real networks. Moreover, we also obtained the
nontrivial clustering coefficient $C$, degree assortativity
coefficient $r$ and degree-strength correlation, depending on a
model parameter $m$. These results are supported by present
empirical evidences. Studying the degree-dependent average
clustering coefficient $C(k)$ and the degree-dependent average
nearest neighbors' degree $k_{nn}(k)$ also provide us with a
better description of the hierarchies and organizational
architecture of weighted networks.
\end{abstract}


\maketitle
\section{Introduction}
In the past few years, there has been a great devotion of
physicists to understand and characterize the underlying
mechanisms of complex networks, e.g. the Internet \cite{Internet},
the WWW \cite{WWW}, the scientific collaboration networks (SCN)
\cite{CN1,CN2,CN3} and world-wide airport networks
(WAN)\cite{air1,air2,air3}. Until now, network researchers have
mainly focused on the topological aspect of graphs, that is,
unweighted networks. Typically, Barab\'asi and Albert proposed a
famous model (BA model) that introduces the linear degree
preferential attachment mechanism to study unweighted growing
networks \cite{BA}. In their model, however, if one considers
other measures like the clustering coefficient then one may
conclude that this model is still insufficient to describe
reality. The hypothesis of a linear attachment rate is empirically
supported by measuring different real networks, but the origin of
the ubiquity of the linear preferential attachment is not clear
yet. Most recently, the availability of more complete empirical
data has allowed scientists to consider the variation of the
weights of links that reflect the physical characteristics of many
real networks. It is well-known that networks are not only
specified by their topology but also by the dynamics of weight
taking place along the links. For instance, the heterogeneity in
the intensity of connections may be very important in
understanding network systems. The amount of traffic
characterizing the connections of communication systems or large
transport infrastructure is fundamental for a full description of
these networks. Take the WAN for example: each given edge weight
$w_{ij}$ (traffic) is the number of available seats on direct
flight connections between the airports $i$ and $j$. In the SCN,
the nodes are identified with authors and the weight depends on
the number of coauthored papers. Obviously, there is a tendency
for a modelling approach to networks that goes beyond the purely
topological point of view. In the light of this need, Alain
Barrat, et al. presented a model (BBV model) that integrates the
topology and weight dynamical evolution to study the growth of
weighted networks \cite{BBV}. Their model yields scale-free
properties of the degree, weight and strength distributions,
controlled by an introduced parameter $\delta$. However, its
weight dynamical evolution is triggered only by newly added
vertices, hardly resulting in satisfying interpretations to the
collaboration networks or the airport systems.

The properties of a graph can be expressed via its adjacency
matrix $a_{ij}$, whose elements take the value 1 if an edge
connects the vertex $i$ to the vertex $j$ and 0 otherwise. The
data contained in the previous data sets permit to go beyond this
topological representation by defining a weighted graph. Weighted
networks are often described by a weighted adjacency matrix
$w_{ij}$ which represents the weight on the edge connecting
vertices $i$ and $j$, with $i,j=1,\ldots,N$, where $N$ is the size
of the network. We will only consider undirected graphs, where the
weights are symmetric ($w_{ij}=w_{ji}$). As confirmed by
measurements, complex networks often exhibit a scale-free degree
distribution $P(k)\thicksim k^{-\gamma}$ with 2$\leq\gamma\leq$3
\cite{air1,air2}. The weight distribution $P(w)$ that any given
edge has weight $w$ is another significant characterization of
weighted networks, and it is found to be heavy tailed, spanning
several orders of magnitude \cite{ref1}. A natural generalization
of connectivity in the case of weighted networks is the vertex
strength described as $s_{i}=\sum_{j\in\Gamma(i)}w_{ij}$, where
the sum runs over the set $\Gamma(i)$ of neighbors of node $i$.
The strength of a vertex integrates the information about its
connectivity and the weights of its links. For instance, the
strength in WAN provides the actual traffic going through a vertex
and is obvious measure of the size and importance of each airport.
For the SCN, the strength is a measure of scientific productivity
since it is equal to the total number of publications of any given
scientist. This quantity is a natural measure of the importance or
centrality of a vertex in the network. Empirical evidence
indicates that in most cases the strength distribution has a fat
tail \cite{air2}, similar to the power law of degree distribution.
Highly correlated with the degree, the strength usually displays
scale-free property $s\thicksim k^{\beta}$ \cite{traffic-driven,
empirical,empirical2,hierarchy}.

The previous models of complex networks always incorporate the
(degree or strength) preferential attachment mechanism, which may
result in scale-free properties. Essentially speaking, this
mechanism just describes interactions between the newly-added node
and the old ones. The fact is that such interactions also exist
between old nodes. This perspective has been practised in the work
of Dorogovtsev and Mendes (DM) \cite{DM} who proposed a class of
undirected and unweighted models where new edges are added between
old sites (internal edges) and existing edges can be removed (edge
removal). On the other hand, we argue that any connection is a
result of mutual affinity and attachment between nodes, while many
network models seem to ignore this point. Traditional models often
present us such an evolution picture: pre-existing nodes are
passively attached by newly adding nodes according to linear
degree (or strength) preferential mechanism. This picture is just
a partial aspect for most complex networks. It is worth remarking
that the creation and reinforcement of internal connections are an
important aspect for understanding real graphs.

In this paper, we shall present a model for weighted networks that
considers the topological evolution under the general mechanism of
mutual selection and attachment between vertices. It can mimic the
reinforcement of internal connections and the evolution of many
infrastructure networks. The diversity of scale-free
characteristics, nontrivial clustering coefficient, assortativity
coefficient and nonlinear strength-degree correlation that have
been empirically observed can be well explained by our microscopic
mechanisms. Moreover, in contrast with previous models where
weights are assigned statically \cite{ref2,ref3} or rearranged
locally \cite{BBV}, we allow weights to be widely updated.

\section{The Mutual Selection Model}
The model starts from an initial configuration of $N_{0}$ vertices
fully connected by links with assigned weight $w_{0}$. The model
is defined on two coupled mechanisms: the topological growth and
the mutual selection dynamics:

(i){\it Topological Growth.} At each time step, a new vertex is
added with $n$ edges connected to $n$ previously existing
vertices, choosing preferentially nodes with large strength; i.e.
a node $i$ is chosen according to the strength preferential
probability:
\begin{equation}
\Pi_{new\rightarrow i}=\frac{s_i}{\sum_ks_k}.
\end{equation}The weight of each new edge is also fixed to $w_{0}$.

(ii){\it Mutual Selection Dynamics}. Every existing node $i$
selects $m$ other pre-existing nodes for possible connection
according to the probability:
\begin{equation}
\Pi_{i\rightarrow j}=\frac{s_j}{\sum_ks_k-s_i}.
\end{equation}Considering the normalization requirement and that
vertices are not permitted to connect themselves, the denominator
of $\Pi_{i\rightarrow j}$ contains the term $-s_i$. If two
unconnected nodes are mutually selected, then an internal
connection is built between them. If two connected nodes are
mutually selected, then their connection is strengthened; i.e.
their edge weight is increased by $w_{0}$. Here the parameter $m$
is the number of candidate vertexes for creating or strengthening
connections. Later, we will see that $m$ also controls the growing
speed of the network's total strength; for instance, the
increasing rate of total information in a communication system.

We argue that connections in most real networks are due to the
mutual selections and attachments between nodes. Take the SCN for
example: the collaborations among scientists require their common
interest and mutual acknowledgements. Unilateral effort does not
promise collaboration. Two scientists that both have strong
scientific potentials (large strengths) and long collaborating
history are more likely to publish papers together during a
certain period. Likewise, for the Movie Actor Collaboration
Networks (MACN), two actors that both have high popularity, if
they co-star, are more probably to boost up the box office. So it
is reasonable to assume that each node is more likely to choose
those nodes with large strengths when building or strengthening
connections. But pre-existing nodes with large strengths will not
be passively attached by nodes with small strength. There are
competition and adaptation in such a complex systems. Both natural
and social networks bear such a property or mechanism during their
evolutions. The above description of our model could
satisfactorily explain the WAN too. The weight here denotes the
relative magnitude of the traffic on a flight connection. At the
beginning of the construction of the airport network, the air
traffic is usually open between metropolises that hold a high
status in both economy and politics. Once a new airline is created
between two airports, it will trigger a more intense traffic
activity depending on the specific nature of the network topology
and on the micro dynamics. With the improvement of national
economy and the expansion of population, the air traffic between
metropolises will increase. Due to their importance, there is an
obvious need for other cities to build new airports to connect the
metropolises. Yet, it is reasonable that the traffic between
metropolises will grow faster than that between other cities, each
of which holds a lower economical and political status and a
smaller population that can afford airplanes. But due to the limit
of energy and resources, each node can only afford a limited
number of connections. Hence facing the vertex pool, they have to
choose. For instance, in the WAN, an airport can not afford the
cost of connecting all the other airports.

The network provides the substrate on which numerous dynamical
processes occur. Technology networks provide large empirical
database that simultaneously captures the topology and the
dynamics taking place on it. For Internet, the information flow
between routers (nodes) can be represented by the corresponding
edge weight. The total information load that each router deals
with can be denoted by the node strength, which also represents
the importance of given router. The increasing information flow as
an internal demand always spurs the expansion of technological
networks. Specifically, the largest contribution to the growth is
given by the emergence of links between already existing nodes.
This clearly points out that the Internet growth is strongly
driven by the need of a redundancy wiring and an increasing need
of available bandwidth for data transmission \cite{empirical}. On
one end, newly-built links (between existing routers) are supposed
to preferentially connect high strength routers, because
otherwise, it would lead to the unnecessary traffic congestion
along indirect paths that connect those high strength nodes.
Naturally, information traffic along existing links between high
strength routers, in general, increases faster than that between
low strength routers. This phenomenon could be reproduced in our
model too. On the other end, new routers preferentially connect to
routers with larger bandwidth and traffic handling capabilities
(the strength driven attachment). This characteristic also exists
in airport system, power grid, and railroad network; and they
could be explained by our mechanisms.

\section{Probability Distributions and Strength Evolution}
The network growth starts from an initial seed of $N_0$ nodes, and
continues with the addition of one node per unit time, until a
size $N$ is reached. Hence the model time is measured with respect
to the number of nodes added to the graph, i.e. $t=N-N_0$, and the
natural time scale of the model dynamics is the network size $N$.
Using the continuous approximation, we can treat $k$, $w$, $s$ and
the time $t$ as continuous variables \cite{Internet, BA}. Then the
edge weight $w_{ij}$ is updated as this evolution equation:
\begin{eqnarray}
\frac{dw_{ij}}{dt}&=&m\frac{s_j}{\sum_ks_k-s_i}\times
m\frac{s_i}{\sum_ks_k-s_j} \nonumber\\
&=&\frac{m^{2}s_is_j}{(\sum_ks_k-s_j)(\sum_ks_k-s_i)}.
\end{eqnarray}
There are two processes that contribute to the increment of
strength $s_i$. One is the creation or reinforcement of internal
connections incident with node $i$, the other is the attachment to
$i$ by newly added node. So the rate equation of strength $i$ can
be written as below:
\begin{eqnarray}
\frac{ds_i}{dt}&=&\sum_j\frac{dw_{ij}}{dt}+n\times\frac{s_i}{\sum_ks_k} \nonumber\\
&\approx&\frac{m^{2}s_i}{\sum_ks_k}\frac{\sum_js_j}{\sum_ks_k}+\frac{ns_i}{\sum_ks_k} \nonumber\\
&=&(m^{2}+n)\frac{s_i}{\sum_ks_k}.
\end{eqnarray}
This equation may be written in a more compact form by noticing
that
\begin{equation}
\sum_{i=1}^ts_i=\int_0^t\sum_{k\in\Lambda}\frac{ds_k}{dt}dt+nt\approx(m^{2}+2n)t,
\end{equation}where $\Lambda$ represents the set of existing
nodes at time step $t$. By plugging this result into the equation
(4), we obtain the following strength dynamical equation
\begin{equation}
\frac{ds_i}{dt}=\frac{m^{2}+n}{m^{2}+2n}\frac{s_i}{t},
\end{equation}which can be readily integrated with initial
conditions $s_i(t=i)=n$, yielding
\begin{equation}
s_i(t)=n(\frac{t}{i})^{\frac{m^{2}+n}{m^{2}+2n}}.
\end{equation}The equation $\sum_is_i\approx(m^{2}+2n)t$ also
indicates that the total strength of the vertices in statistical
sense is uniformly increased with the size of network. As one see,
the growing speed of the network's total strength load is mainly
determined by the model parameter $m$.

The knowledge of the time-evolution of the various quantities
allows us to compute their statistical properties. Indeed, the
time $t_i=t$ at which the node $i$ enters the network is uniformly
distributed in [0,t] and the degree probability distribution can
be written as
\begin{equation}
P(s,t)=\frac{1}{t+N_0}\int_{0}^{t}\delta(s-s_i(t))dt_i,
\end{equation}
where $\delta(x)$ is the Dirac delta function. Using equation
$s_i(t)\sim(t/i)^{\theta})$ obtained from Eq. (7), one obtains in
the infinite size limit $t\rightarrow\infty$ the distribution
$P(s)\sim s^{\alpha}$ with $\alpha=1+1/\theta$:
\begin{equation}
\alpha=2+n/(m^{2}+n).
\end{equation}Obviously, when $m=0$ the model is topologically
equivalent to the BA network and the value $\alpha=3$ is
recovered. For larger values of $m$, the distribution is gradually
broader with $\alpha\rightarrow2$ when $m\rightarrow\infty$.

\begin{figure}
\scalebox{0.80}[0.75]{\includegraphics{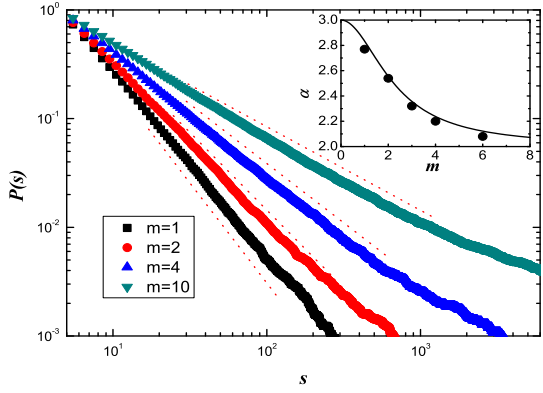}}
\caption{\label{fig:epsart} Probability distribution $P(s)$. Data
are consistent with a power-law behavior $s^{-\alpha}$. In the
inset we give the value of $\alpha$ obtained by data fitting
(filled circles), together with the analytical expression
$\alpha=2+n/(m^{2}+n)=2+5/(m^{2}+5)$(line). The data are averaged
over 10 independent runs of network size N=5000.}
\end{figure}

\begin{figure}
\scalebox{0.80}[0.75]{\includegraphics{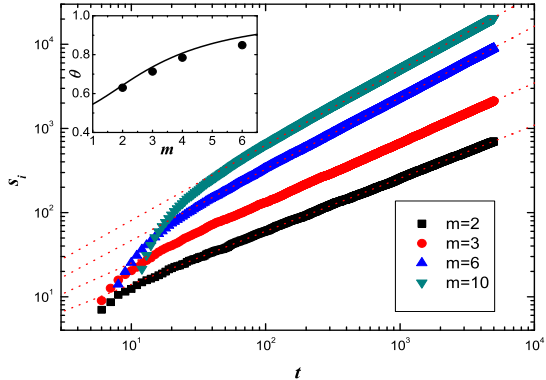}}
\caption{\label{fig:epsart} Evolution of strength of vertices
during the growth of network for various of $m$. In the inset we
give the value of $\theta$ obtained by data fitting (filled
circles), together with the analytical expression
$\theta=\frac{m^{2}+n}{m^{2}+2n}=\frac{m^{2}+5}{m^{2}+10}$(line).}
\end{figure}

\begin{figure}
\scalebox{0.80}[0.75]{\includegraphics{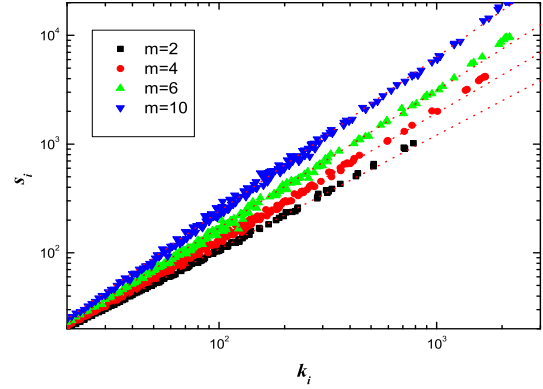}}
\caption{\label{fig:epsart} Strength $s_i$ versus $k_i$ for
different $m$ (log-log scale). Linear data fitting gives slope
1.04, 1.16, 1.26 and 1.41 (from bottom to top), demonstrating the
correlation of $s\sim k^{\beta}$.}
\end{figure}

\begin{figure}
\scalebox{0.80}[0.75]{\includegraphics{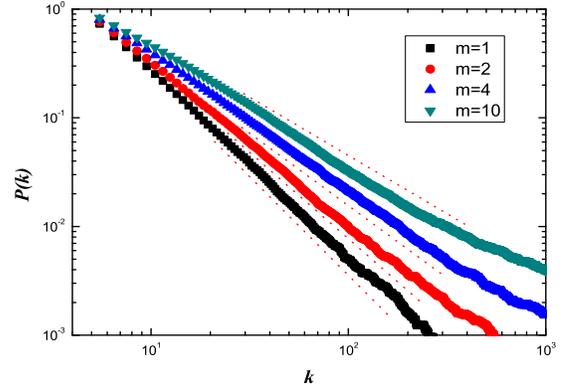}}
\caption{\label{fig:epsart} Probability distribution of the
degrees $P(k)\sim k^{-\gamma}$ for different $m$. The data are
averaged over 10 independent runs of network size N=5000.}
\end{figure}

\begin{figure}
\scalebox{0.80}[0.75]{\includegraphics{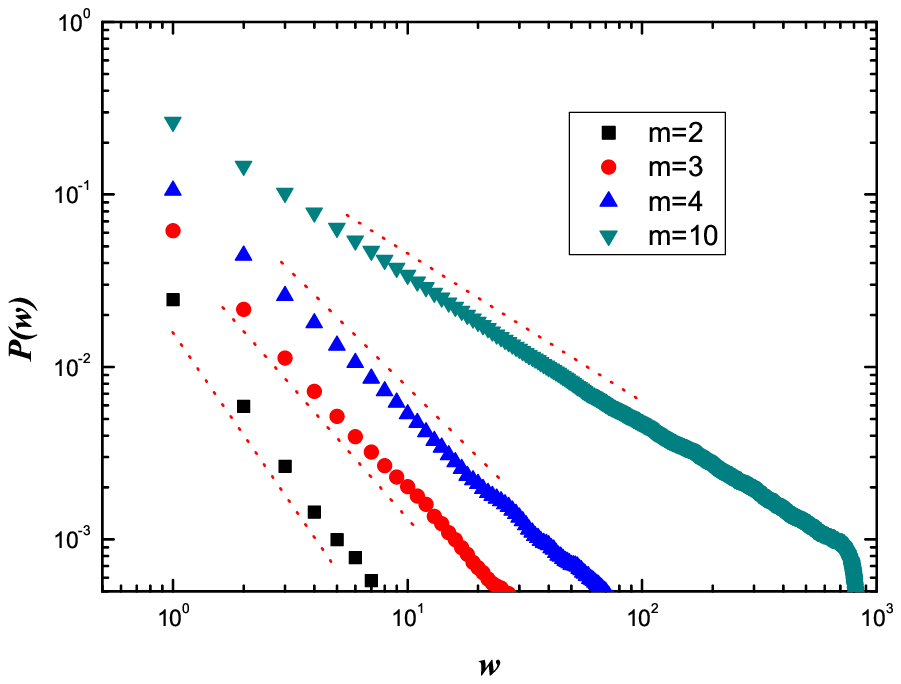}}
\caption{\label{fig:epsart} Probability distribution of the
weights $P(w)\sim w^{-\eta}$ for various $m$. The data are
averaged over 10 independent runs of network size N=5000.}
\end{figure}

We performed numerical simulations of networks generated by
choosing different values of $m$ and fixing $n=5$ and $w_{0}=1$.
Considering that every vertex strength can at most increase by $m$
from internal connections and a newly added node can at most
connect with $n$ existing nodes, it is easy to see that the
initial network configuration must satisfy $N_0=\max(m+1,n)$. So
for example, if $m=10$, then $N_0=11$. We have checked that the
scale-free properties of our model networks are independent of the
initial conditions. Numerical simulations are consistent with our
theoretical predictions, verifying again the reliability of our
present results. Fig. 1 gives the probability distribution
$P(s)\sim s^{\alpha}$, which is in excellent agreement with the
theoretical predictions. In Fig. 2 we show the behavior of the
vertices' strength versus time for different values of $m$,
recovering the behavior predicted by analytical methods. We also
report the average strength $s_i$ of vertices with degree $k_i$,
which displays a nontrivial power-law behavior $s\sim k^{\beta}$
as confirmed by empirical measurement. Unlike BBV networks (where
$\beta=1$), the exponent $\beta$ here varies with the parameter
$m$ in a nontrivial way as shown in Fig. 3. The nontrivial $s\sim
k^{\beta}$ correlation demonstrates the significant part of weight
increment along existing edges. More importantly, one could check
the scale-free property of degree distribution ($P(k)\sim
k^{-\gamma}$) by combining $s\sim k^{\beta}$ with $P(s)\sim
s^{-\alpha}$. Considering the conservation of probability
\begin{equation}
\int_0^{\infty}P(k)dk=\int_0^{\infty}P(s)ds,
\end{equation}we can readily calculate the exponent $\gamma$:
\begin{eqnarray}
P(k)=P(s)\frac{ds}{dk}=s^{-\alpha}\beta k^{\beta-1}=\beta
k^{-(\beta(\alpha-1)+1)},
\end{eqnarray}
giving $\gamma=\beta(\alpha-1)+1$. The scale-free properties of
degree and weight obtained from simulations are presented in Fig.
4 and Fig. 5, respectively. The simulation consistence of
scale-free properties indicates that our model can indeed produce
power-law distributions of degree, weight and strength. In this
case, the numerical simulations of the model reproduce the
behaviors predicted by the analytical calculations.

\section{Clustering and Correlation}
Many real networks in nature and society share two generic
properties: they are scale-free and they display a high degree of
clustering. Along with the general vertices hierarchy imposed by
the scale-free strength distribution, complex networks show an
architecture imposed by the structural and administrative
organization of these systems, which is mathematically encoded in
the various correlations existing among the properties of
different vertices. For this reason, a set of topological and
weighed quantities are customarily studied in order to uncover the
network architecture. A first and widely used quantity is given by
the {\it clustering} of vertices. The clustering of a vertex $i$
is defined as
\begin{equation}
c_i=\frac{1}{k_i(k_i-1)}\sum_{j,h}a_{ij}a_{ih}a_{jh},
\end{equation}
and measures the local cohesiveness of the network in the
neighborhood of the vertex. Indeed, it yields the fraction of
inter-connected neighbors of a given vertex. The average over all
vertices gives the network {\it clustering coefficient} which
describes the statistics of the density of connected triples.
Further information can be gathered by inspecting the average
clustering coefficient $C(j)$ restricted to classes of vertices
with degree $k$:
\begin{equation}
C(k)=\frac{1}{NP(k)}\sum_{i,k_i=k}c_i.
\end{equation}
In many networks, the average clustering coefficient $C(k)$
exhibits a highly nontrivial behavior with a power-law decay as a
function of $k$\cite{hierarchy}, indicating that low-degree nodes
generally belong to well interconnected communities (high
clustering coefficient) while high-degree sites are linked to many
nodes that may belong to different groups which are not directly
connected (small clustering coefficient). This is generally the
signature of a nontrivial architecture in which hubs (high degree
vertices) play a distinct role in the network. Numerical
simulations indicate that for large $m$, the clustering
coefficient $C(N)$ is almost independent of $N$ (as we can see in
Fig. 6), which agrees with the finding in several real networks
\cite{BA}. Generally, when the network size $N$ is larger than
5000, the clustering coefficient is nearly stable. So most
computer runs are assigned with 5000. Still, it is worth remarking
that for the BA networks, $C(N)$ is nearly zero, far from the
practical nets that exhibit a variety of small-world properties.
In the present model, however, clustering coefficient $C$ is
fortunately found to be a function of $m$ (see Fig. 7), also
supported by empirical data in a broad range. Finally, the
clustering coefficient $C(k)$ depending on connectivity $k$ for
increasing $m$ is also interesting and shown in Fig. 8. For
clarity, we add the dashed line with slope $-1$ in the log-log
scale. This simulation results are supported by recent empirical
measurements in many real networks. For the convenience of
comparison with Fig. 8, we use two figures from Ref. [17] as our
Fig. 9, from which one can see the agreements of theory and
experiment are quite excellent. Empirical evidence generally
supports the simulations of clustering-degree correlation. Though
some previous models\cite{mm1,mm2,mm3} can generate the power-law
decay of the clustering-degree correlation, none of them as far as
we know can produce the flat head as found in real graphs. This is
a special property that our model successfully behaves.

\begin{figure}
\scalebox{0.80}[0.75]{\includegraphics{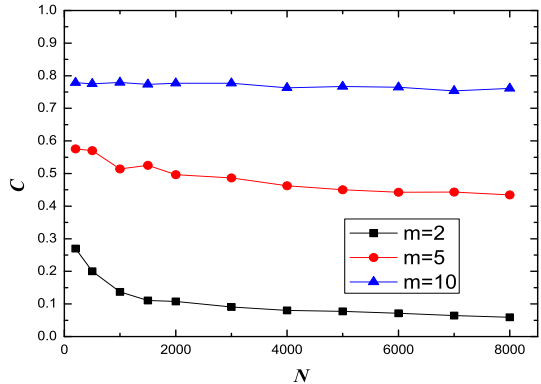}}
\caption{\label{fig:epsart} The evolution of clustering
coefficient (or $C$ versus $N$) which converges soon.}
\end{figure}

\begin{figure}
\scalebox{0.80}[0.75]{\includegraphics{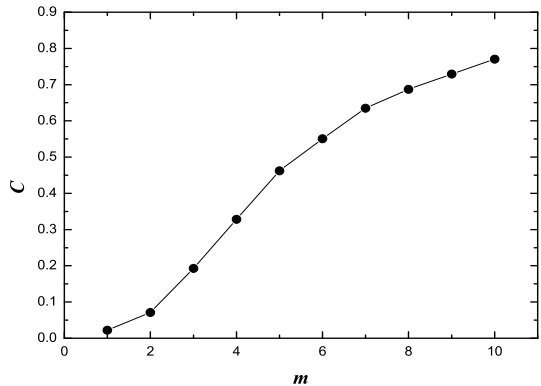}}
\caption{\label{fig:epsart} Clustering coefficient $C$ depending
on the parameter $m$.}
\end{figure}

\begin{figure}
\scalebox{0.80}[0.75]{\includegraphics{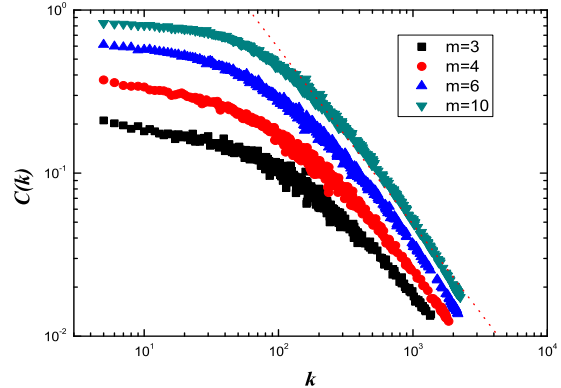}}
\caption{\label{fig:epsart} The clustering coefficient $C(k)$
depending on connectivity $k$ for increasing $m$. For comparison,
the dashed line has slope -1 in the log-log scale.}
\end{figure}

\begin{figure}
\scalebox{0.80}[0.75]{\includegraphics{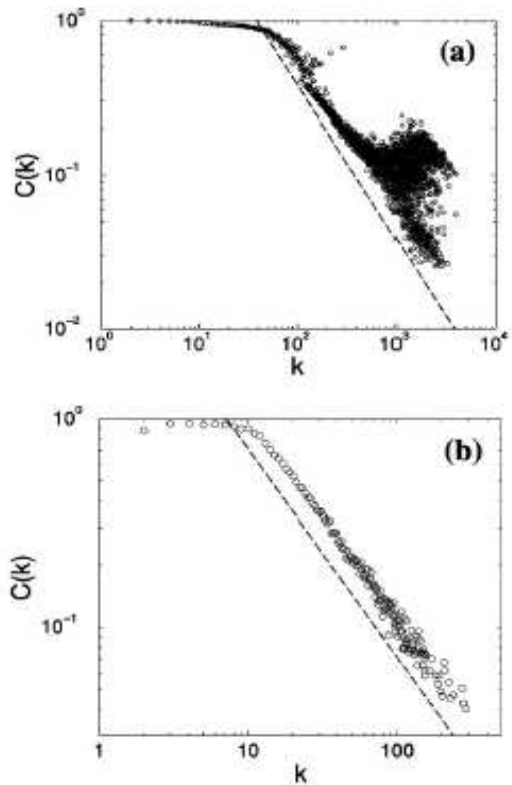}}
\caption{\label{fig:epsart}The scaling of $C(k)$ with $k$ for two
real networks \cite{hierarchy}: (a)Actor network, two actors being
connected if they acted in the same movie according to the
www.IMDB.com database. (b)The semantic web, connecting two English
words if they are listed as synonyms in the Merriam Webster
dictionary. The dashed line in each figure has slope -1.}
\end{figure}

Another important source of information is the correlations of the
degree of neighboring vertices. The {\it average nearest neighbor
degree} is proposed to measure these correlations
\begin{equation}
k_{nn,i}=\frac{1}{k_i}\sum_{j}a_{ij}k_j.
\end{equation}
Once averaged over classes of vertices with connectivity $k$, the
average nearest neighbor degree can be expressed as
\begin{equation}
k_{nn}(k)=\sum_{k'}P(k'|k),
\end{equation}
providing a probe on the degree correlation function. If degrees
of neighboring vertices are uncorrelated, $P(k'|k)$ is only a
function of $k'$ and thus $k_{nn}(k)$ is a constant. When
correlations are present, two main classes of possible
correlations have been identified: {\it assortative} behavior if
$k_{nn}(k)$ increases with $k$, which indicates that large degree
vertices are preferentially connected with other large degree
vertices, and {\it disassortative} if $k_{nn}(k)$ decreases with
$k$. The above quantities provide clear signals of a structural
organization of networks in which different degree classes show
different properties in the local connectivity structure. In the
light of this measure, we also perform computer simulations to
test the $k_{nn}(k)-k$ correlation, as shown in Fig. 10. As
$k_{nn}(k)$ decreases with $k$, one may find that our model can
best illustrate disassortative networks in reality, that mainly
is, technological networks (e.g. Internet, WAN) and biological
networks (e.g. Protein Folding Networks). As for the social
networks, connections between people may be assortative by
language or by race. Newman proposed some simpler measures to
describe these types of mixing, which we call assortativity
coefficients \cite{mixing}. Almost all the social networks studied
show positive assortativity coefficients while all others,
including technological and biological networks, show negative
coefficients. It is not clear if this is a universal property; the
origin of this difference is not understood either. In our views,
it represents a feature that should be addressed in each network
type individually. In the following, we use the formula proposed
by Newman in Ref. \cite{mixing},
\begin{equation}
r=\frac{M^{-1}\sum_ij_ik_i-[M^{-1}\sum_i\frac{1}{2}(j_i+k_i)]^{2}}
{M^{-1}\sum_i\frac{1}{2}(j_i^{2}+k_i^{2})-[M^{-1}\sum_i\frac{1}{2}(j_i+k_i)]^{2}},
\end{equation}
where $j_i$, $k_i$ are the degrees of vertices at the ends of the
$i$th edges, with $i=1,...,M$ ($M$ is the total number of edges in
the observed graph). We calculate the degree assortativity
coefficient (or degree-degree correlation) $r$ of the graphs
generated by our model. For large $N$ (e.g. $N>5000$), the
degree-degree correlation $r$ is almost independent of the network
size (see Fig. 11). Simulations of $r$ depending on $m$ are given
in Fig. 12 and supported by empirical measurements for
disassortative networks \cite{mixing}.

\begin{figure}
\scalebox{0.90}[0.75]{\includegraphics{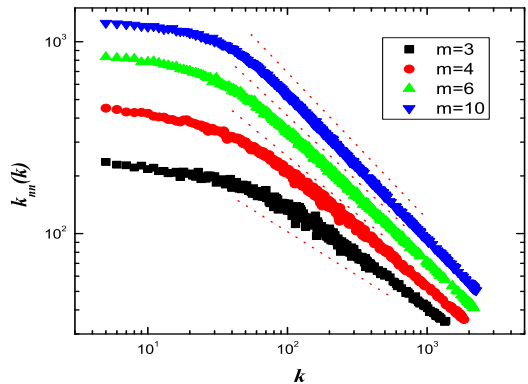}}
\caption{\label{fig:epsart} Average connectivity $k_{nn}(k)$ of
the nearest neighbors of a node depending on its connectivity $k$
for different $m$.}
\end{figure}

\begin{figure}
\scalebox{0.80}[0.75]{\includegraphics{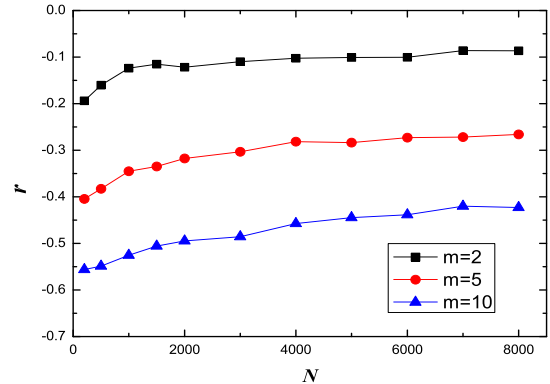}}
\caption{\label{fig:epsart} Degree-degree correlation $r$
depending on $N$. The evolution of $r$ converges soon.}
\end{figure}

\begin{figure}
\scalebox{0.80}[0.75]{\includegraphics{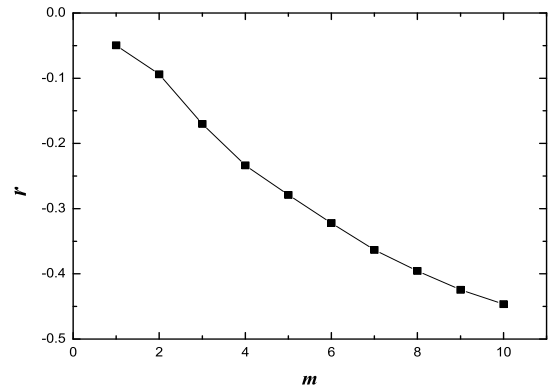}}
\caption{\label{fig:epsart} Degree-degree correlation $r$
depending on $m$.}
\end{figure}

\section{Conclusion and outlook}
In sum, integrating the mutual selection mechanism between nodes
and the growth of strength preferential attachment, our network
model provides a wide variety of scale-free behaviors, tunable
clustering coefficient and nontrivial (degree-degree and
strength-degree) correlations, just depending on the parameter $m$
which governs the total weight growth. All the results of network
properties are found supported by various empirical data.
Interestingly and specially, studying the degree-dependent average
clustering coefficient $C(k)$ and the degree-dependent average
nearest neighbors' degree $k_{nn}(k)$ also provide us with a
better description of the hierarchies and organizational
architecture of weighted networks. As far as our knowledge, there
at present appears no model that could generate so many properties
which are in broad agreement with empirical data. Thus, our model
may be very beneficial for future understanding or characterizing
real networks. Though our model can just produce disassortative
networks (most suitable for technological and biological ones),
which means one of its limitations, we always expect some model
versions or variations that generate weighted networks with
assortative property. Due to the apparent simplicity of our model
and the variety of tunable results, we believe that some of its
extensions will probably help address (e.g. social) networks.
Therefore, our present model for all practical purposes will
demonstrate its applications in future weighted network research.

\begin{acknowledgments}

This work has been partially supported by the National Natural
Science Foundation of China under Grant No. 70471033, 10472116 and
No.70271070, the Specialized Research Fund for the Doctoral
Program of Higher Education (SRFDP No.20020358009), and the
Foundation for Graduate Students of University of Science and
Technology of China under Grant No. KD200408.

\end{acknowledgments}

\end{document}